\documentclass[useAMS,usenatbib,usegraphicx]{mn2e}

\def\hunit {\, {\rm km \, s^{-1} \,Mpc^{-1} } } 
 
\def\zeq {z_{\rm eq}} 
\def\omrad{\omega_{\rm rad}} 
\def\neff{ N_{\rm eff}} 
\def\xrad{ X_{\rm rad}} 

\def\simlt{\mathrel{\lower0.6ex\hbox{$\buildrel {\textstyle <} 
  \over {\scriptstyle \sim}$}}}

\def\simgt{\mathrel{\lower0.6ex\hbox{$\buildrel {\textstyle >}
 \over {\scriptstyle \sim}$}}}

\def\newtwo { }

\title[On measuring the absolute scale of BAOs]
{On measuring the absolute scale of baryon acoustic oscillations}

\author[Will Sutherland] 
{Will Sutherland$^{1}$\thanks{E-mail: w.j.sutherland@qmul.ac.uk} 
\\
$^{1}$School of Physics and Astronomy, Queen Mary University of London, 
  Mile End Road, London E1 4NS 
} 

\begin{document}

\date{MNRAS - Accepted 2012 July 5. Received 2012 July 5; in original 
 form 2012 May 3 }

\pagerange{\pageref{firstpage}--\pageref{lastpage}} \pubyear{2012}

\maketitle

\label{firstpage}

\begin{abstract}

 The baryon acoustic oscillation (BAO) feature in the distribution of galaxies 
  provides a fundamental standard ruler which is widely used 
  to constrain cosmological parameters. 
 In most analyses, the comoving length of the ruler 
  is inferred from a combination of CMB observations and theory. 
 However, this inferred length
  may be biased by various
  non-standard effects in early universe physics; this can
  lead to biased inferences of cosmological parameters such as
  $H_0$, $\Omega_m$ and $w$, so it would be valuable to  
  measure the {\em absolute} BAO length by combining 
  a galaxy redshift survey and a suitable direct
   low--$z$ distance measurement.
 One obstacle is that low-redshift BAO surveys
 mainly constrain the ratio $r_S / D_V(z)$, where $D_V$ is a dilation scale
  which is not directly observable by standard candles. 
 Here, we find a new approximation 
 $D_V(z) \simeq \frac{3}{4} D_L(\frac{4}{3} z) (1+ \frac{4}{3} z)^{-1}
   (1 - 0.02455 \, z^3 + 0.0105 \, z^4 ) $ 
   which connects $D_V$ to the standard luminosity distance 
   $D_L$ at a somewhat higher redshift; this 
  is shown to be very accurate (relative error $< 0.2$ percent) 
  for all WMAP-compatible Friedmann models at $z < 0.4$, 
  with very weak dependence 
  on cosmological parameters $H_0$, $\Omega_m$, $\Omega_k$, $w$.  
  This provides a route to measure the {absolute} BAO length using 
  only observations at $z \simlt 0.3$, 
  including type-Ia supernovae, and potentially
   future $H_0$-free physical distance 
  indicators such as gravitational lenses or
  gravitational wave standard sirens.
 This would provide a {\em zero-parameter} check of 
  the standard cosmology at $10^3 \simlt z \simlt 10^5$,
  and can constrain 
  the number of relativistic species $\neff$ with fewer degeneracies 
 than the CMB.

\end{abstract}

\begin{keywords}
cosmology:  large-scale structure of Universe -- 
 cosmic microwave background -- distance scale.  
\end{keywords}

\section{Introduction}

The detection of baryon acoustic oscillations (BAOs) in the 
 large-scale distribution of galaxies in both the SDSS \citep{eis05} and 
 2dFGRS \citep{cole05} redshift surveys was a triumph for the 
 standard cosmological model;
 the BAO feature \citep{peeb-yu70, be84, eh98, mwp99} 
 is essentially created by closely related physics
  to the acoustic peaks in the 
 cosmic microwave background (CMB) temperature power spectrum.  
 Therefore, the observed BAO feature supports the standard
 cosmology in several independent ways: 
  the existence of the feature supports
 the basic gravitational instability paradigm for structure
  formation;  the relative
  weakness of the BAO feature supports the 
  $\sim 1:5$ ratio of baryons to dark matter, since a baryon-dominated
  universe would have a BAO feature much stronger than observed; 
  and the observed length-scale of the
   feature in redshift space is consistent with
  the concordance $\Lambda$CDM model derived from the
   CMB and other observations, with
   $\Omega_m \approx 0.27$ and $H_0 \approx 70 \hunit$
  \citep{komatsu11}.   

 Recently, there have been several new independent 
 measurements of the BAO feature in galaxy redshift surveys, e.g. from 
 SDSS-DR8 \citep{perc10}, WiggleZ \citep{blake11}, 6dFGRS \citep{beutler11}, 
 and an angular measurement from SDSS-DR9 \citep{seo12}, 
 which are all consistent with
 the concordance $\Lambda$CDM model at the few-percent level. 

 The BAO feature is probably the best-understood standard
  ruler in the moderate-redshift Universe, and therefore 
 in conjunction with CMB observations it offers 
  great power for constraining cosmological parameters including
  dark energy \citep{wein12}. 
 A number of theoretical and numerical 
  studies \citep{seo08,seo10} have concluded that
 the comoving length-scale of the BAO feature evolves by
  $\sim 0.5\,$ percent between the CMB era and $z \sim 0.3$ 
  due to non-linear growth of structure, but this 
  shift can be corrected to the $0.1\,$ percent level using 
  reconstruction methods \citep{pad12}.  
 Therefore, there are several very ambitious redshift surveys including
 the ongoing WiggleZ \citep{blake11}, BOSS \citep{white11} and HETDEX, 
  the recently approved ESA Euclid space
  mission \citep{euclid-red}, 
   and the proposed BigBOSS and WFIRST, which plan to survey
  $\sim 1 - 50$ million galaxy redshifts over huge volumes, 
  to give sub-percent measurements
 of the BAO feature at various redshifts $0.2 \simlt z \simlt 2.5$.  

 However, one drawback of the BAO feature is that the
  comoving length $r_s(z_d)$ is calibrated
 at $z > 1000$ using a combination of CMB observations and theory; 
 this leaves us vulnerable to systematic errors from 
  possible unknown new physics at early times (see \S\ref{sec:bao} 
  for discussion). 
 Low-redshift measurements of the BAO scale essentially
  measure a ratio of $r_s$ relative to some distance which is itself
  dependent on cosmological parameters $H_0, \Omega_m, w$ etc. Therefore,
  in a joint fit to CMB+BAO data, a wrong assumption in 
  the CMB measurement of $r_s$ may be masked
   by biased values of cosmological parameters,  
 i.e. a ``precisely wrong'' outcome (see \S\ref{sec:neff} for 
  an example). 
   
In this letter we present a new and useful approximation which can
  accurately anchor the absolute BAO lengthscale using only 
  low redshift measurements at $z \simlt 0.3$, therefore providing
 a clean zero-parameter test of the standard early-universe cosmology,
  in particular the density of relativistic particles.   

The plan of the paper is as follows: in \S~\ref{sec:bao}
 we briefly review the main features of BAO observations, 
 then in \S~\ref{sec:dvdl} we present the new approximation 
 for the dilation scale $D_V(z)$. In \S~\ref{sec:neff}
 we review the effects of non-standard radiation density, 
 and in \S~\ref{sec:dist} we discuss 
 potential observational issues for measuring the absolute 
 BAO length.  We summarise our conclusions in \S~\ref{sec:conc}.  

Throughout the paper we use the standard notation that
 $\Omega_i$ is the present-day density of species $i$ relative to the
 critical density; and the physical density $\omega_i \equiv \Omega_i h^2$, 
 with $h \equiv H_0 / (100 \hunit)$. 
 Our default model is flat $\Lambda$CDM with $\Omega_m = 0.27$; 
  in other cases, $w$ is the dark energy
  equation of state, $\Omega_{tot} = \Omega_m + \Omega_\Lambda$ 
  is the total density parameter, and $\Omega_k \equiv 1 - \Omega_{tot}$.  
 

\section{Observations of the BAO length} 
\label{sec:bao} 

 The BAO feature appears as a single hump in the galaxy correlation
 function $\xi(r)$, or equivalently a series of decaying wiggles
  in the power spectrum (see \cite{esw07} for a 
  clear exposition, and \cite{bh10} for a recent review).  
 In the following, 
   we denote $r_s$ to be the comoving length scale of the BAO feature
  in a galaxy redshift survey, 
  while $r_s(z_d)$ is the comoving sound horizon
  at the baryon drag epoch $z_d \approx 1020$, 
  which is conventionally defined as the fitting formula 
   Eq.~4 of \citet{eh98}. 
  These lengths are not quite identical due to 
  evolution of perturbations after the drag epoch and non-linear
  growth of structure, 
 but the shifts are predicted to be below the 0.6 percent level and well
  correctable from theory \citep{esss07,seo08,seo10}; 
  therefore, measuring the BAO feature
  at low redshift provides a very robust link to the sound 
  horizon in the CMB era.  

 In the small angle approximation and
  assuming we observe a redshift shell which is thin compared
  with its mean redshift, 
 the angular size of the BAO feature is
   $\Delta \theta(z) = r_s / (1+z) D_A(z)$, 
 where $D_A(z)$ is the usual proper angular-diameter distance
  to redshift $z$; 
 and the difference in redshift along one BAO length in the
  line-of-sight direction is 
  $\Delta z_{/\!/} (z) = r_s \, H(z) / c $ \citep{bg03,se03}
 
 In practice, current galaxy redshift surveys are not quite
 large enough to distinguish the BAO feature separately
 in angular and radial directions, so the current measurements
  constrain a spherically-averaged length scale; 
 the most model-independent quantity derived from these
  observations is $r_s / D_V(z)$,  
  where $D_V$ is called the dilation scale and is 
  defined by \citet{eis05} as 
\begin{equation}
 \label{eq:dv} 
  D_V(z) \equiv \left[ (1+z)^2 D_A^2(z) \frac{c z}{H(z)} \right]^{1/3} 
\end{equation} 
 This is essentially a geometric mean of two transverse and one
  radial directions. 
 
  Measuring the BAO feature from a galaxy redshift survey
  requires a mapping from observed galaxy positions and redshifts to
  galaxy pair separations in comoving coordinates, which is itself dependent
  on cosmological parameters including $H_0, \, \Omega_m, \, w$ etc.  
  Therefore, extracting the value of $r_s$ from a galaxy redshift
  survey is slightly theory-dependent;  
  but the above dimensionless ratios $\Delta\theta(z)$, $\Delta z_{/\!/}(z)$ 
  and $r_s / D_V(z)$ 
  are essentially theory-independent, since to first order
  any change in the reference cosmology produces an equal
   shift in the fitted length $r_s$.  

 As above, the comoving length $r_s(z_d)$ is defined
 as the sound horizon at the baryon drag epoch
  $z_d \approx 1020$ \citep{eh98}. 
    Adopting standard early-universe assumptions,
  $r_s(z_d)$ depends only on the densities of matter, baryons and 
 radiation; the latter is very well constrained by the CMB temperature
  (assuming standard neutrino content), 
  hence $r_s$ depends only on $\omega_m$ and $\omega_b$, which in turn
 are well constrained by the heights and morphology of the
 first three CMB peaks. Fits
  from the WMAP7 data alone \citep{komatsu11} 
  give $r_s(z_d) \approx 153 \, {\rm Mpc}$ with approximately
  1.3~percent precision,  
  and forthcoming data from the Planck mission \citep{planck-miss} 
  is expected to improve this prediction of $r_s$ to 
  $\approx 0.3\,$ percent precision.  
 
 We note that the CMB derivation  
  of $r_s(z_d)$ does not rely on assuming a flat universe or
 details of dark energy,  
  since the observed CMB peak heights 
 constrain $\omega_m$ and $\omega_b$ well without assuming flatness. 
  However, the inference of $r_s(z_d)$ 
  does rely on many simple but weakly tested  
  assumptions about the $z > 1000$ universe, including 
\begin{enumerate} 
\item Standard GR, 
\item Standard radiation content 
  (photons plus an effective number $\neff \approx 3.04$ light neutrinos), 
\item  Standard recombination history, including negligible variation in 
  fundamental constants. 
\item  Negligible early dark energy, 
\item  Negligible contribution of isocurvature fluctuations, 
\item  The primordial power spectrum is smooth and almost a power-law. 
 \item  Densities of matter and radiation scale as the 
  standard powers of scale factor; e.g. negligible late-decaying 
   particles at $z \simlt 10^6$ etc.  
\end{enumerate}  

 If one or more of the above assumptions are wrong, this can bias
  the value of $r_s$ deduced from the CMB fits, 
  and in turn this will generally result in biased
  inferences about other cosmological parameters (especially
  $H_0$) when doing joint fits to CMB and BAO data.  
  Possible violation of (ii) above was analysed by 
  \citet{ew04}, and is discussed later in \S~\ref{sec:neff}; 
   for some other non-standard cases,
  see for example \citet{lin-rob} for early dark energy, 
 \citet{menegoni} for varying $\alpha$, and \citet{zunckel} for
  the effect of isocurvature fluctuations. 

  For the above reasons, 
   measuring the {\em absolute} BAO length scale at low redshift 
   forms a powerful consistency test of the assumptions underlying 
  standard cosmology at $z > 1000$.  

 The well-known route to this is to measure the transverse BAO
  scale at some effective redshift which gives 
  $r_s/(1+z)D_A(z)$, and also use a combination of standard candles
  (e.g. SNe Ia)  
  and the local Hubble constant $H_0$ to measure 
 the usual luminosity distance, $D_L(z)$, to the same redshift. 
  Combined with the standard distance-duality 
  relation $D_L(z) = (1+z)^2 D_A(z)$, 
  this can give a theory-independent absolute measurement 
   of the BAO length. 
 However, one disadvantage of the above is that it requires a
 BAO survey of sufficiently large volume to separate the transverse and
  radial BAO scales, and reaching sufficient cosmic volume requires a survey
  at significant redshift $z \simgt 0.3$; 
 in turn, this means that type-Ia SNe are likely
 the only distance indicators bright enough to be useful for measuring 
  $D_L(z)$, and there is a non-negligible time baseline 
 over which supernova evolution may bias the measurements of $D_L(z)$. 

 As a complement to the above, 
  it would be valuable to calibrate $r_s$ using
 BAO measurements of $r_s/D_V(z)$  
 at lower redshifts $0.1 \simlt z \simlt 0.25$, combined
  with an accurate calibration of $D_V(z)$ from distance indicators.  
 Although BAO surveys at lower redshift suffer from 
   increased cosmic variance due to the limited available volume, 
  there are several compensating benefits: 
 there is a shorter time baseline for possible evolution 
   of SNe properties; 
 the SNe are brighter and more readily observable in the
  rest-frame near-IR; 
 and low redshift offers better prospects for using 
  alternative distance indicators
 such as gravitational lens time-delays, 
 and potentially gravitational-wave
 standard sirens \citep{einstein-tel}; and 
 finally we avoid the complication of separating 
 the radial and angular BAO scales in the analysis.  

 However, this low-$z$ route requires an absolute 
  measurement of $D_V(z)$ rather than $D_A(z)$, 
  which is slightly more challenging; from Eq.~\ref{eq:dv},
   a measurement of $D_L(z)$ tells
  us $D_V(z)$ apart from an unknown factor of $H(z)^{1/3}$; this is
  helpful due to the 1/3 power of $H$, 
   but is not good enough for percent-level precision. 
   At $z \rightarrow 0$, $D_V(z) \rightarrow cz /H_0$ as with
   all cosmological distances; however, 
  there is insufficient volume to measure the BAO feature 
   at $z \simlt 0.05$, while beyond this cosmological distance 
    effects cannot be ignored.   
  For a concordance $\Lambda$CDM model at an example $z = 0.2$, the 
  crude approximation $D_V(z) \sim cz / H_0$ 
   is 6~percent too large, while the approximation 
  $D_V(z) \approx D_L(z)/(1+z)$ is too large by 1.6~percent; these 
  approximations are clearly not good enough for precision cosmology.  

 Since $D_V(z)$ is directly related to the comoving volume element 
  per unit redshift, via 
 \begin{equation} 
 \label{eq:vol} 
  {dV_c \over dA \, dz} =  { c (1+z)^2 \, D_A^2(z) \over H(z) } 
 = {1 \over z} D_V^3(z) 
\end{equation} 
  where $dV_c$ is comoving volume and $dA$ is solid angle, 
 we could measure $D_V$ directly if we could observe  
  a population of ``standard counters'' of known comoving number
 density. Unfortunately, our limited understanding of
  galaxy evolution implies that there is little
  hope of finding standard counters good enough for
 a percent-level measurement of $D_V$. 
 Alternatively, a direct measurement of $H(z)$ is possible using
  differential ages of red galaxies (e.g. \citealt{stern10}), but again
  it may be very challenging to reach few-percent 
  absolute accuracy with this method.  

 In the next Section, we show a new alternative
  route for obtaining accurate calibration of $D_V$: we
  find a much better approximation for $D_V$, which relates
  $D_V(z)$ to the observable $D_L$ at a slightly higher redshift, 
  specifically $\frac{4}{3}z$.   

\section{ A route to measuring $D_V$ } 
\label{sec:dvdl} 

\subsection{ Relation between $D_V$ and $D_L$ } 

Here we find an an accurate approximation relating
 the dilation length $D_V(z)$ to the observable luminosity 
 distance $D_L$ at a slightly higher redshift. 

We first define as usual the scale factor $a \equiv (1+z)^{-1}$ with
  $a_0 = 1$,  
 and the time-dependent Hubble parameter $H(z) = \dot{a}/a$ where
  dot denotes time derivative.  
We also have the usual expression for comoving radial distance,
\begin{equation} 
 D_R(z_1) = c \int_0^{z_1} {1 \over H(z)} \, dz 
\end{equation} 
In the Appendix of \citet{suth12}, we found a good approximation 
at moderate redshift 
 \begin{equation} 
 D_R(z) \approx {c z \over H(\frac{z}{2})} \ , 
\end{equation} 
This approximation was derived using a Taylor-series 
  expansion of $1/H(z)$ around redshift $z/2$; this results in the 
  first derivative $(1/H)'$ cancelling so there is no error of
 order $z^2$,  and uses the convenient fact that
  the second derivative $(1/H)''$ has a zero-crossing at $z \sim 0.3$
 in a concordance $\Lambda$CDM model, so the error of order $z^3$ is 
 small at moderate redshift.  
 In a flat universe this leads to 
\begin{equation} 
\label{eq:dlapp} 
  D_L(z) \approx (1+z) { c z \over H(\frac{z}{2}) } \ . 
\end{equation} 
This is still fairly accurate even for weakly-curved 
 models, since the multiplicative change in $D_L$ for
 a non-flat model is of order $1 + \Omega_k z^2 / 6$ 
  for fixed expansion history; for plausible values of
 $\vert \Omega_k \vert < 0.05$, this is a very small 
  effect at $z \simlt 0.3$.  

\begin{figure}  
\includegraphics[angle=-90, width=9cm]{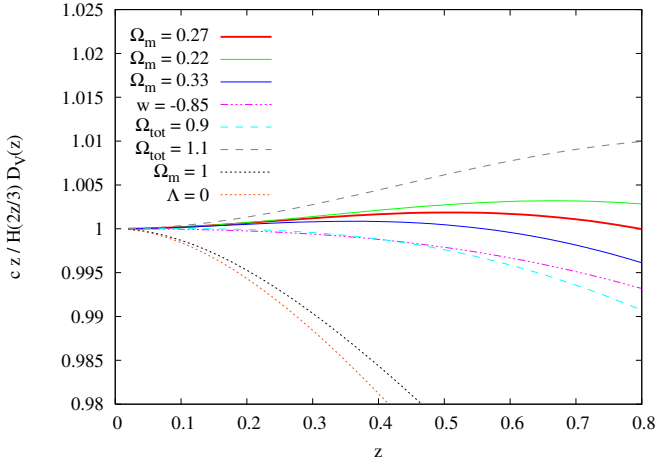} 
\caption{
 This figure shows the relative accuracy of
 approximation~\ref{eq:dvh} 
  for various cosmological models. 
 The solid lines show flat $\Lambda$CDM models
  with $\Omega_m = 0.22, \; 0.27, \;  0.33$ {\newtwo (top to bottom)}. 
 The dashed lines show non-flat $\Lambda$CDM with $\Omega_{tot} = 0.90$
  (lower) and $1.1$ (upper). The dash-dot line shows flat $w$CDM with 
  $w = -0.85$. 
 The dotted lines show $\Omega_m = 1$ (upper), and open 
  $\Omega_m = 0.27, \, \Omega_\Lambda = 0$ (lower). 
 }  
\label{fig:dvh} 
\end{figure}

In \citet{suth12} we also found a good approximation for $D_V(z)$
 at moderate redshift,  which is 
\begin{equation}
\label{eq:dvh} 
  D_V(z) \approx {c z \over H(\frac{2}{3} z) } 
\end{equation} 
Both approximations (\ref{eq:dlapp}) and (\ref{eq:dvh})
 are accurate to $\le 0.4\,$ percent for $z < 0.5$ for models reasonably
 close to standard $\Lambda$CDM; the error in approximation (\ref{eq:dvh})
 is shown in Figure~\ref{fig:dvh} 
 for some example models.  (This 
 will be useful below in \S~\ref{sec:omm}). 

Both Eqs.~\ref{eq:dlapp} and \ref{eq:dvh} involve $H(z)$ 
 at slightly different redshifts; however, it is clear from the above
 that if we consider a BAO measurement at effective redshift $z_1$, 
  then $D_V(z_1)$ is closely related to $H(2z_1/3)$,
 while if we consider $D_L(4 z_1/3)$ this is also related
 to $H(2 z_1/3)$; we can therefore combine approximations \ref{eq:dlapp}
 and \ref{eq:dvh}    
  to cancel the unknown $H(2 z_1 /3)$, which gives the approximation 
\begin{equation} 
\label{eq:dvapp1} 
  D_V(z) \simeq {3 \over 4} \, D_L\left(\frac{4}{3} z\right) 
  \left(1+ \frac{4}{3}z \right)^{-1} \ . 
\end{equation} 

\begin{figure*}  
\includegraphics[width=10cm, angle=-90]{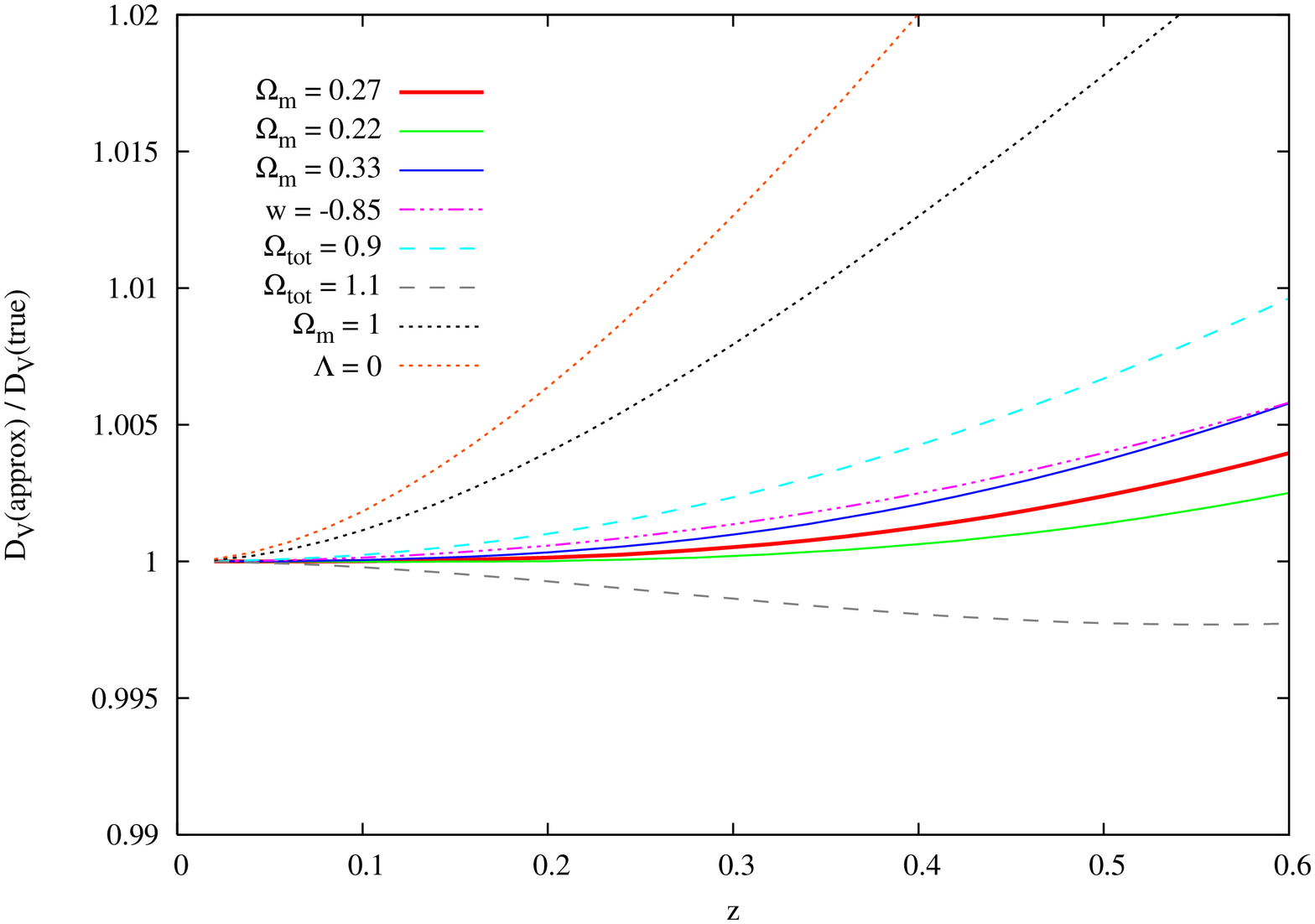} 
\caption{
 This figure shows the relative accuracy of
 approximation~\ref{eq:dvapp1} 
  for the same models as Figure 1. 
 The three solid lines show flat $\Lambda$CDM models
  with $\Omega_m = 0.22, \; 0.27, \;  0.33$ (bottom to top). 
 The dashed lines show non-flat $\Lambda$CDM with $\Omega_{tot} = 0.90$
  (upper) and $1.1$ (lower). The dot-dash line shows flat $w$CDM with 
  $w = -0.85$. 
 The dotted lines show $\Omega_m = 1$ Einstein-de Sitter (lower) and
  open $\Omega_m = 0.27, \Omega_\Lambda = 0$ (upper). 
 }  
\label{fig:dvapp1} 

\includegraphics[width=10cm, angle=-90]{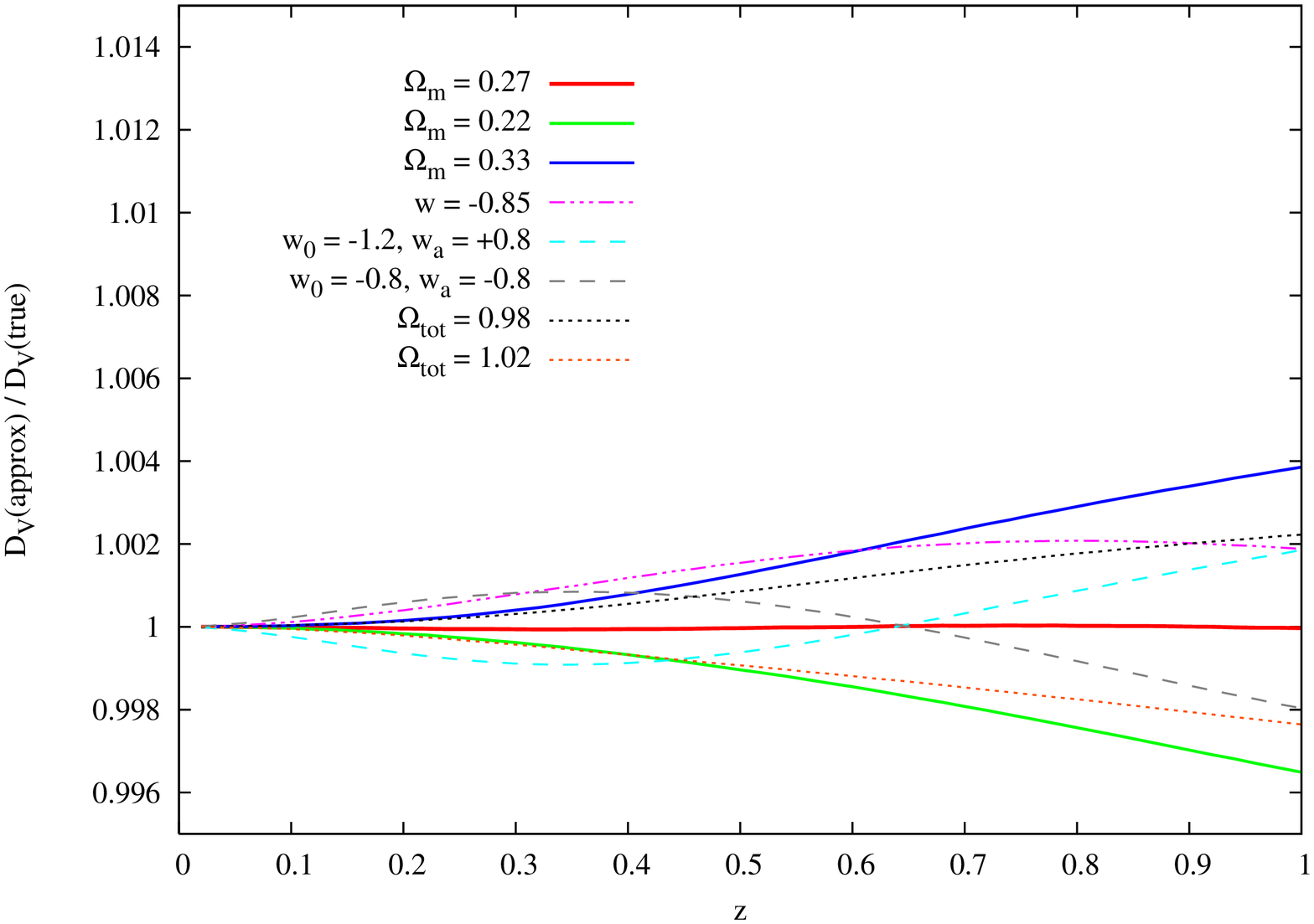} 
\caption{
 This figure shows the relative accuracy of
 approximation~\ref{eq:dvapp2} 
  for various cosmological models, roughly consistent with WMAP.  
 The solid lines show flat $\Lambda$CDM models with $\Omega_m = 0.22, \; 0.27,
  \;  0.33$ (bottom to top). 
 The dash--dotted line shows flat $w$CDM with constant $w = -0.85$; 
  the dashed lines show evolving-$w$ models 
  with values $(w_0,w_a) = (-1.2, +0.8)$ and $(-0.8, -0.8)$ respectively. 
 The dotted lines show non-flat $\Lambda$CDM with 
  $\Omega_{tot} = 0.98$ (upper) and $1.02$ (lower). 
 (Note the axis scales are different from Figure 2). 
 }  
\label{fig:dvapp2} 
\end{figure*}

We now explore the error in approximation \ref{eq:dvapp1}:   
  Figure~\ref{fig:dvapp1} shows the ratio 
 of the RHS of Eq.~\ref{eq:dvapp1} to the exact $D_V(z)$ 
  for various example cosmological models. 
 Unless otherwise specified, we take $\Omega_m = 0.27$ for 
  each model. 
 Specifically, Figure~\ref{fig:dvapp1} shows three flat $\Lambda$CDM models 
 with $\Omega_m = 0.22,\, 0.27,\,  0.33$; 
  one flat $w$CDM model with $w = -0.85$;  
 and two non-flat $\Lambda$CDM models with $\Omega_{tot}$ = 0.9 and
  1.1 respectively; finally,  an Einstein-de Sitter $\Omega_m = 1$ 
  model, and a zero dark energy model with 
 $\Omega_m = 0.27, \, \Omega_{\Lambda} = 0$. 
  (These latter two models are well known to be grossly
  inconsistent with CMB and other measurements, but are included for
  comparison purposes). 
 
It is clear from Figure~\ref{fig:dvapp1} that
 approximation~\ref{eq:dvapp1} is surprisingly accurate: 
 for the three $\Lambda$CDM models the error is less than
  0.2 percent at $z < 0.4$, and the $w$CDM model is only slightly
 worse. 
 The two oCDM models are still quite good, with error $< 0.4$~percent 
  at $z < 0.4$;  
  this is considerably better than the medium-term 
  expected errors $\sim 1\,$ percent on the BAO ratio, 
  and also current upper limits 
  on $\vert \Omega_k \vert $ are significantly
 tighter than 0.1; more realistic values 
 $\vert \Omega_k \vert \sim 0.02$ give rise to
   minimal error in (\ref{eq:dvapp1}).  
 Therefore for any WMAP-allowed model, 
  the error in approximation (\ref{eq:dvapp1}) 
  is several times smaller than the medium-term precision
  on BAO observables. 
 
The approximation \ref{eq:dvapp1} becomes significantly worse for  
 the Einstein-de Sitter and open zero-$\Lambda$ models, 
  with errors respectively +1.25~percent and +2.0~percent at $z = 0.4$; 
  however, even these give sub-percent error at $z \le 0.25$. 

 In fact, Eq.~(\ref{eq:dvapp1}) is exact at all $z$ for a de Sitter model
  with $\Omega_m = 0, \,\Omega_\Lambda = 1$, while its accuracy 
 degrades rather slowly with increasing $\Omega_m$ and/or curvature; thus,
 for near-flat and accelerating models favoured by current data, 
   it is remarkably accurate. 
An explanation of this property in terms of Taylor series is given
 in Appendix A: this shows that (\ref{eq:dvapp1}) is exact to 
 second order in $z$ independent of all cosmological parameters; 
 while at third order, there is a fortunate coincidence that 
  for deceleration parameter $q_0 \simlt -0.4$ and small curvature, 
 the difference in the $z^3$ coefficients is also small. 
 This makes Eq.~\ref{eq:dvapp1} accurate at $z \simlt 0.4$ for all 
  near-flat accelerating models, with little dependence on precise
  values of $\Omega_m$, $\Omega_k$, $w$ etc.   
 {\newtwo (Note that all results in the main body of the paper
 use the numerical integrals for $D_V$ and $D_L$; the Taylor series
 in Appendix A are only provided as an aid to intuition). }  

  We also see from Figure~\ref{fig:dvapp1} that
 approximation~(\ref{eq:dvapp1}) is a slight overestimate 
  of the exact $D_V(z)$ for all the 
 flat and open models shown; only the closed model ($\Omega_{tot} = 1.1$) 
 gives an underestimate.  
 Since (\ref{eq:dvapp1}) gives a slight overestimate of 
  the exact $D_V$ for all the 
  plausible models fairly close to $\Lambda$CDM, 
  we can get a modest but useful improvement by removing this bias, 
  by multiplying by a polynomial in $z$ 
  chosen to give a good fit for concordance $\Lambda$CDM; 
 we find an excellent fit with small terms in $z^3$ and $z^4$, 
  specifically 
\begin{equation} 
\label{eq:dvapp2} 
  D_V(z) \simeq {3 \over 4} \, D_L\left(\frac{4}{3} z\right) 
  \left(1+ \frac{4}{3}z \right)^{-1}  
 \left(1 - 0.0245 \,z^3 + 0.0105 \, z^4 \right) 
  \ . 
\end{equation} 

 By construction, this approximation is excellent for the
  concordance model, with relative error $< 0.02$ percent at $z \le 1$.   
 For other plausible models, 
  the resulting ratio (RHS of \ref{eq:dvapp2}) / (exact $D_V(z)$) 
   is shown in Figure~\ref{fig:dvapp2}; 
  in this Figure we have used a smaller 
  range of $\Omega_k$ for the non-flat models, 
   and added two models with time-varying $w$ with the
  common parametrisation $w(a) = w_0 + w_a (1-a)$, to give a set of  
   models roughly spanning the $2\sigma$ allowed range from 
  current data.   

 It is clear from Figure~\ref{fig:dvapp2} that approximation 
  \ref{eq:dvapp2} is
 very accurate in the WMAP-allowed neighbourhood of $\Lambda$CDM,  
  including generous variations of $\Omega_m$, 
  modest curvature and $w \ne -1$. 
 For all the models shown the relative error 
  is smaller than $(z / 200)$ at $z < 1$, thus 0.2 percent 
  error at $z \le 0.4$. 
 This error is substantially smaller than
   the cosmic variance in BAO measurements, and
 the expected accuracy in next-decade absolute 
  distance measurements, so is almost negligible for practical
  measurements of $r_s$.  
 
 This means that a direct measurement of $D_L(4z/3)$ can immediately
 predict $D_V(z)$ with very little dependence on cosmological
 parameters $H_0, \,\Omega_m$,  $\Omega_k, \, w$.   Multiplying 
  this by a BAO measurement of $r_s/D_V(z)$ from a galaxy redshift survey 
 thus measures $r_s$ in comoving Mpc, based entirely on low-redshift
  data.  

 This can then be compared with a CMB-only prediction of $r_s(z_d)$ 
 for a {\em zero parameter} test of our early 
 universe assumptions: if the local $r_s$ measured
 from BAOs and $D_L$ as above is not consistent 
  with the $r_s$ inferred from the CMB, something
 is definitely wrong with one or more measurements, 
 or the early-universe assumptions or the FRW framework; 
  tuning of the late-time cosmological parameters $\Omega_m, \Omega_k, w$ 
   within the $3\sigma$ WMAP-allowed ranges cannot 
  significantly help. 
 Conversely if a bottom-up measurement of $r_s$ 
  does agree at the $\sim 1-2\,$ percent level
  with the CMB prediction, this would provide simple and compelling  
  support for the standard set of early-universe assumptions.  

We next discuss some questions in both the CMB and
 local methods for measuring $r_s$.

\section{Non-standard radiation density} 
\label{sec:neff} 

Here we consider the effect of non-standard radiation density,
 which is an important degeneracy for all CMB-determined 
  length scales. This is moderately well-known, first
 analysed for the BAO case by \citet{ew04}, 
 and also in e.g. \citet{debernardis}, but we give a slightly
  different and hopefully more intuitive explanation 
 compared to previous work. 

\subsection{Definitions of radiation density} 
\label{sec:neff1} 

 The value of $r_s(z_d)$ is given from the WMAP data alone 
  as 153 Mpc with 2~percent precision, just assuming standard radiation
 content but no assumptions about flatness or dark energy.  
 However, if the assumption of standard radiation content
 is dropped, the precision degrades radically to $> 10\%$ 
 \citep{komatsu11}. The reason is 
 mainly the strong degeneracy between matter density $\omega_m$ and 
 radiation density $\omrad$ in the
  CMB fits:  the first three CMB peaks determine
  the redshift of matter/radiation equality $\zeq$ well, with 
  $1 + \zeq \equiv \omega_m / \omrad \approx 3200 \pm 140$,
  but converting to the physical matter density $\omega_m$ then relies on
  an assumption of the total radiation density.  
 
The radiation density is conventionally parametrised by an effective
 number of neutrino species $\neff$ in the CMB era, defined via 
\begin{equation}  
 \omrad \equiv \omega_\gamma \left[ 1 + \frac{7}{8} \left( {4 \over 11} 
 \right )^{4/3} \neff \right] 
\end{equation}  
 (Here for non-negligible neutrino mass, $\omrad$ is
  not the present-day radiation density, but the value at 
  $\zeq$ rescaled by $(1+\zeq)^{-4}$. We set 
 $\omega_m = \omega_c + \omega_b$ to include dark matter
 and baryons only, excluding any low-redshift contribution
  from neutrino mass). 

Most analyses assume a standard value very close to 
  $\neff = 3.046$ effective neutrino species \citep{mangano05} 
  which gives $\omrad = 1.6918 \,\omega_\gamma$; and
  the photon density $\omega_\gamma = (40440)^{-1}$ is 
  set by the very accurate CMB temperature, $T_0 = 2.7255 \, {\rm K}$.  
 However, we note that there already some hints of a higher value
 of $\neff$ from e.g. \cite{keis11}; these are not yet decisive, 
 but are very interesting. 


For general $\neff$, the above can be rearranged into 
\begin{equation} 
\label{eq:wm} 
 \omega_m = 0.1339 \, \left( \frac{1 + \zeq}{3201} \right) \, 
  \left[ \,1 + 0.134 (\neff - 3.046) \right] 
\end{equation}  

\subsection{The $\neff$/scale degeneracy} 
\label{sec:degen} 

Here we review in more detail the effect of non-standard $\neff$
 on cosmological parameter estimates, and show essentially
 that this creates a degeneracy in overall scale factor which
 affects all cosmic distances, times and densities, 
 but has very little effect on dimensionless ratios.  

It is helpful to rearrange the expression for the 
 sound horizon (e.g. Eq.6 of \citet{eh98}) in terms of 
  $\zeq$, $\omrad$ and $\omega_b$ as the input parameters, which gives  
\begin{eqnarray} 
\label{eq:rsound} 
 r_s(z_d) & = & 2998 \,{\rm Mpc} \frac{2}{\sqrt 3} 
  \, \omrad^{-1/2} 
  \, (1+z_{eq})^{-1} \, R_{eq}^{-1/2} \nonumber \\
   & & \times \, \ln \left( { 
 \sqrt{1 + R_d} + \sqrt{R_d + R_{eq}} \over 1 + \sqrt{R_{eq}} 
  } \right) 
\end{eqnarray} 
 where $R(z) \approx 30330 \, \omega_b / (1+z)$ 
 is the baryon/photon momentum density ratio, and $R_d$, $R_{eq}$
  are the values at $z_d$, $\zeq$ respectively. 
 This shows that if we vary $\omrad$ while holding $\zeq$, $z_d$, and 
 $\omega_b$  all fixed, 
  the sound horizon scales simply $\propto \, \omrad^{-1/2}$.
 In more detail, the WMAP best-fit values show small 
   changes in $z_d$ and $\omega_b$ with varying $\neff$, 
 see Sec~4.7 of \citet{komatsu11} for details:  
   however, the consequential shifts in $r_s(z_d)$ are 
  some $10\times$ smaller 
  than the dominant $\omrad^{-1/2}$ shift, so we ignore those 
  for simplicity. 
 
 Next for illustration we take a specific example of two models, an arbitrary 
  model A with 
 $\neff = 3.046$ and parameters assumed a good fit to WMAP, 
  and a model B with one extra neutrino species (or equivalent
  in dark radiation), thus $\neff = 4.046$ but the same $\zeq$.   
  Thus, model B has both $\omrad$ and $\omega_m$ larger by 13.4~percent,  
   while the sound horizon in B is smaller by a 
   factor close to $(1.134)^{-1/2} = 0.939$.   
  This would be severely discrepant with the observed position of the
 CMB acoustic peaks via the acoustic scale $\ell_*$,   
   if either the distance to last-scattering 
 $D_A(z_*)$ or $H_0$ were held fixed.  
 
  However, if we also choose model B to have
 increased $\omega_{DE}, \omega_k$ 
   by the same factor of 1.134 above, then via
 \begin{equation}   
   h^2 = \omega_m + \omega_{DE} + \omega_k \quad , 
 \end{equation}
   model B has $H_0$ increased by 6.5~percent 
  while all of $\Omega_m$, $\Omega_{DE}$, $\Omega_{rad}$, $\Omega_k$ 
   are identical in models A and B. 
  Since the expansion function $E(z) \equiv H(z)/H_0$ depends only on the 
   upper--case $\Omega$ values above, $E(z)$ 
    remains unchanged at all redshifts;   
  so {\bf all} cosmic distance$(z)$ and $t(z)$
  functions are reduced by $\approx 6\%$ in model B relative to A, 
   but distance {\em ratios} between 
  any two redshifts (related to BAO and SNIa observables and the CMB 
  acoustic angle) remain unchanged. 
 (We note here that $\omega_b$ and $\omega_\gamma$ are assumed unchanged
   between models A and B, so 
   $\Omega_b$ and $\Omega_b/\Omega_m$ are reduced by 13 percent in model B, 
   but these do not appear separately in the Friedmann equation. 
  The implied value of $\sigma_8$ will also be slightly different
  for model B as shown by WMAP, but
  we do not consider that here). 

 What is happening here is simple: apart from $\omega_b$, 
  WMAP mainly constrains dimensionless
  quantities: especially $\zeq$, the acoustic scale $\ell_*$ at 
  last scattering $z_*$, and the shift parameter
 $\cal{R}$. Also, BAO measurements are intrinsically dimensionless 
 ratios such as $r_s/D_V(z)$, while supernova measurements 
 anchored to the local Hubble flow also give dimensionless ratios,
  essentially $H_0 \, D_L(z)/c$ or $D_L(z) / D_L(z = 0.03)$
 (while $H_0$ is degenerate with the standardised SN luminosity).
  All these above provide precision 
   measurements of the uppercase $\Omega$ values and $w$ with no overall
  scale needed. 
  
 But, there are three dimensionful quantities (lengths, times and densities,
 or combinations of these)
 in homogeneous cosmology; while there are {\em two} inter-relations:
 distances and times are related by the
 known $c$, and the Friedmann equation relates densities to 
  expansion rate, via $G$.  
 This implies that even excellent knowledge of all those dimensionless
  ratios above is not sufficient to solve for any 
  dimensionful quantity; but adding a measurement
 of {\em any one} cosmological length, time, or absolute density of matter,
 radiation or dark energy (in SI units or equivalent) would be
   sufficient to constrain all the others.  
  Usually, this dimensionful quantity is (implicitly) set
  by assuming $\neff \approx 3.046$, which fixes $\omrad$ and thus
  all the other scales:   
  but if this assumption is dropped, then WMAP+BAO+SNe observations
  leave us short by one dimensionful
  quantity, and the $\neff$ vs $H_0$ degeneracy appears.  

 Given the above, it is convenient to define the scaled radiation
  density as 
\begin{equation}
\label{eq:xrad} 
  \xrad \equiv \omrad / 1.692 \, \omega_\gamma \, = 1 + 0.134 (\neff - 3.046) 
\end{equation} 
 so that the standard value is 1; and also to choose a fundamental
  parameter set including
\begin{equation} 
\label{eq:pars} 
  \Omega_m \ ; \ \zeq \ ; \xrad \ ; \ \omega_b \   
\end{equation} 
  plus optional parameters $\Omega_k, \, w$ defaulting to $0, -1$; 
  as usual $\Omega_{DE} = 1 - \Omega_m - \Omega_k$.   
 This set couples very naturally to the observables, and   
  turns both $\omega_m$ and $H_0$ 
  into derived parameters,  via 
\begin{eqnarray} 
\label{eq:hzeq} 
\omrad & = & \xrad / 23904 \nonumber \\ 
\omega_m & = & (1+\zeq) \xrad / 23904 \nonumber \\ 
   h & \equiv & \sqrt(\omega_m / \Omega_m) 
\end{eqnarray}  

 Currently, the observational uncertainty on $\xrad$ is substantially larger
  than on the other major parameters: the central value
 depends somewhat on choice of datasets,
 with some datasets favouring $\neff \approx 4$ (e.g. \citealt{keis11}), 
  while others prefer the standard 
  $\neff \approx 3$ (e.g. \citealt{mangano11}). 
 There is broad agreement that 
  $2 \le \neff \le 5$, which maps to $0.86 \le \xrad \le 1.27$. 
 We find from WMAP results that if we allow $\xrad \ne 1$, then 
  current cosmological
 measurements mainly constrain the combinations
 $\omega_m \approx (0.135 \pm 0.005) \,\xrad $, 
  $ r_s \approx (153 \pm 2) / \sqrt{\xrad} \, {\rm Mpc}$, 
  $H_0  \approx (70 \pm 1.5) \sqrt{\xrad} \,\hunit$ and $t_0 \approx 
  (13.75 \pm 0.1) / \sqrt{\xrad} \, {\rm Gyr}$; 
  all these dimensionful observables
 have error bars dominated by the 
  uncertainty in $\xrad$,  while most dimensionless ratios are
  nearly uncorrelated with $\xrad$ (the main exceptions are
  $n_s$ and $\sigma_8$, which both show small positive correlations with
  $\xrad$).  This simple scaling accurately reproduces the degeneracy
    track of $t_0$ vs $\neff$ shown by \citet{debernardis}. 

 (We emphasise an important distinction that $h$ and $\omega_i$ 
  {\em do not} count as dimensionless in this discussion; 
 they are clearly pure numbers, but they represent dimensionful quantities
  rescaled by an arbitrary choice of $H_0 = 100 \hunit$ 
  and a fiducial density $\rho_{fid} = 1.878 \times 10^{-26} 
  {\rm \, kg \, m^{-3}}$.  The true
 dimensionless quantities such as $\zeq$, $\Omega_m$, $H_0 D_A(z)/c$,
  $H_0 r_s/c$, $\ell_*$, $H_0 t_0$, etc,   
  have values independent of any system of units).  

 To break the above degeneracy, 
 it is sufficient to get an accurate measurement of {\em any one} dimensionful 
 observable such as $H_0$, $t_0$, $r_s$ or an absolute distance $D_L(z)$
  to any redshift. 
 (A purely local measurement of $\omrad$, $\omega_m$ or $\omega_{DE}$
   would also suffice, but appears impossible). Other possibilities
 include the CMB damping tail (see below) at $\ell > 1000$, 
 which brings in the {\newtwo Silk damping length}  
  as a new dimensionful quantity which has different scaling
  with $\xrad$.   

 Of the various dimensionful parameters above, 
  $H_0$ is clearly the most familiar from history, 
 but to constrain $\neff$ it is actually preferable to measure $r_s$:     
 because $r_s \sqrt{\xrad}$ is well constrained by CMB data alone, 
  independent of late-time dark energy and curvature.  
 In contrast, $H_0 / \sqrt{\xrad}$ is well constrained by WMAP+BAO data 
  {\em if we assume} flatness and $w = -1$, but the constraints become 
  significantly worse if we allow curvature and/or arbitrary 
   dark energy evolution. 
  Therefore, measuring the absolute length $r_s$ is the best route to 
   probe $\neff$ and the early Universe; 
  while combining CMB data with an accurate local $H_0$ measurement mainly
  constrains one degenerate combination of $\neff$, $w$ and $\Omega_k$. 

  To illustrate this more clearly, the concordance model values of 
   $H_0 \approx 70.5 \hunit$
  and $r_s \approx 153 \, {\rm Mpc}$ shift to $\approx 75 \hunit$ and 
   144 Mpc respectively if we assume $\Lambda$CDM with 
    $\neff \approx 4.0$. If the latter is the actual cosmology,  
   an observed lower bound $H_0 \ge 73 \hunit$ could be fitted
   with any of $\neff \approx 4$, or $\neff \approx 3$ with
  $w < -1$ and/or open curvature;  but a direct upper bound 
   $r_s \le 148 \,{\rm Mpc}$ would exclude all of the 
   currently allowed range for $\neff \approx 3$, and decisively
   {\em require} extra radiation or some other new physics at $z > 1000$. 

  Also, it is helpful to compare 
   the $\neff$/scale degeneracy
   to the better-known geometrical degeneracy affecting parameter 
  fitting from the CMB alone \citep{eb99}.
  Although both degeneracies affect $H_0$, 
  the geometrical degeneracy involves holding fixed physical densities
  of both matter and radiation (thus fixed $r_s$), 
  while trading off two of 
   $\Omega_m$, $\Omega_{k}$, $w$ so as to maintain a fixed angular distance to
  last-scattering $D_A(z_*)$; 
   this is well broken by BAO ratios, as we see below.  
   The $\neff$/scale degeneracy above also 
   holds $\zeq$ fixed, but rescales 
    densities and distances by $\xrad$ and $1 / \sqrt{\xrad}$ respectively;
   here both $D_A(z_*)$ and $r_s(z_*)$ shift by a common factor.    
  This $\neff$/scale degeneracy is {\em not} broken by BAO distance-ratios, 
   but is broken with an {\em absolute} BAO length measurement. 
  Therefore, these two degeneracies are ``orthogonal'' concerning
   $r_s$, but get mixed in $H_0$,  which explains why $r_s$ is a cleaner
   test of the early Universe. 

\subsection{An easy route to $\Omega_m$} 
\label{sec:omm} 
Here we find a strikingly simple
  route to $\Omega_m$, accurate to better than 1 percent: 
 first it is convenient to define
\begin{equation} 
\label{eq:epsv} 
  1 + \epsilon_V (z;\Omega_m,\Omega_k,w) \equiv 
  {c z \over H(\frac{2}{3} z) \,  D_V(z) } 
\end{equation} 
 so the function $\epsilon_V$ is defined to be the (small) correction to 
   approximation (\ref{eq:dvh}), as shown in Figure~\ref{fig:dvh}.  
 Then, taking an observed BAO ratio $r_s / D_V(z)$, 
  substituting Eq.~\ref{eq:rsound}  
  and using 
  $h / \sqrt{\omrad} \equiv \sqrt{(1+\zeq) /\Omega_m } $,  we obtain  
\begin{eqnarray} 
 \label{eq:eom} 
 {z \, r_s \over D_V(z)} & = & { r_s H(\frac{2}{3}z) \over c} 
  (1 + \epsilon_V) \nonumber \\ 
   & = & (1 + \epsilon_V) 
  { E(\frac{2}{3}z) \over \sqrt{\Omega_m} }  {2 \over \sqrt 3} 
  (1 + \zeq)^{-1/2} R_{eq}^{-1/2}  \nonumber \\
  & & \times \ln \left( { \sqrt{1 + R_d} + \sqrt{R_d + R_{eq}} 
  \over 1 + \sqrt{R_{eq}} } \right)  \ .
\end{eqnarray}  
 This is exact apart from non-linear shifts of $r_s$. 
  All the terms above are clearly dimensionless:  
 both $H_0$ and $\omrad$ have cancelled, and there is only 
  a small implicit dependence on $\omrad$ via very small changes in $z_d$. 

 The last three factors on the RHS above are well constrained given 
  only $\zeq$ and $\omega_b$ from WMAP, which are almost independent of 
   dark energy, curvature or radiation density.  
 Adopting $\omega_b = 0.0225$, the RHS above is very well
  approximated by 
\begin{equation} 
\label{eq:omez} 
  {z \, r_s \over D_V(z)} \simeq 0.01868 \, (1+\epsilon_V) 
 { E(\frac{2}{3}z) \over \sqrt{\Omega_m} }  
  \left( { 1 + \zeq \over 3201 } \right)^{0.25} \ ; 
\end{equation}  
  with the uncertainty due to $\omega_b$ below 0.4 percent.
 
 A precise moderate-redshift BAO measurement from SDSS is given by 
  \citet{pad12} as $D_V(z = 0.35) / r_s = 8.88 \pm 0.17$; 
 thus the LHS above is $0.0394 \pm 0.0008$. 
 This, together with $\zeq \approx 3200 \pm 130$ 
  and neglecting the sub-percent $\epsilon_V$ term 
 gives $E(0.233) / \sqrt{\Omega_m} = 2.11 \pm 0.05$; simply 
 squaring this and rearranging gives a linear relation (for $w = -1$) of 
 $\Omega_m = (0.280 + 0.145 \,\Omega_k) (1 \pm 0.05)$, in excellent
 agreement with the full likelihood results.  

 There is a common rule-of-thumb that ``CMB measures
 $\omega_m$ and the BAOs measure $H_0$''; we see from the above
 that this is {\em only} valid {\newtwo assuming}
  the standard $\neff \simeq 3.0$. 
 For general radiation density, 
  the CMB is really measuring $\zeq$, not $\omega_m$:  
  adding a low-redshift
  BAO measurement then measures primarily $\Omega_m$, with a small
   sensitivity to $\Omega_k$ and $w$ creeping in via the $E(2z/3)$ term.  
  Combining $\zeq$ and $\Omega_m$ gives us a value for $H_0 / \sqrt{\xrad}$ 
  from Eq.~(\ref{eq:hzeq}),  again with mild dependence on $\Omega_k, w$,  
  but does not give an absolute scale.  

 (The additional information from 
  the large-scale bend in the galaxy power spectrum is discussed in 
  Appendix~B). 

\section{Distance and sound horizon measurements} 
\label{sec:dist}  

 Here we discuss some considerations on observational issues and the
  realistic precision available for 
 measurements of the absolute BAO length, both locally and
 from future CMB measurements.  

\subsection{Distance ladder measurements} 
\label{sec:ladder} 

Given a measurement of $r_s/D_V(z)$ from BAOs in a redshift survey, 
  we need an absolute measurement of $D_L(4z/3)$ to apply Eq.~\ref{eq:dvapp2} 
  and obtain an absolute measurement of $r_s$ independent
 of the CMB. 
  The most obvious route to measure $D_L(4z/3)$ is to combine  
 a local distance-ladder measurement of $H_0$ with
  a large sample of type-Ia supernovae centred at redshift 
  near $4z/3$ to measure $D_L(4z/3)$; along with 
 approximation~\ref{eq:dvapp2} this provides
 a direct calibration of $D_V(z)$ and thus $r_s$. 

 Doing this at fairly low redshift has several advantages:
 firstly, it is observationally much cheaper to accumulate a large
 sample of supernovae at $z \simlt 0.3$ compared with $z \simgt 0.5$,
 and such a sample should arise naturally from the ongoing PanSTARRS
 Medium Deep Survey \citep{pan-starrs} and the Dark Energy Survey 
  \citep{des-sn}.   
 Also, lower redshift provides a smaller lever-arm for possible
 time evolution of the mean supernova brightness, minimising
 systematic errors. Given an overabundance of supernovae (e.g. several
  hundred in the relevant redshift bin), 
  one can afford to subdivide the sample by lightcurve stretch, host 
 galaxy type etc, to provide consistency checks. 

Also, there is growing evidence that SNe Ia are closest to 
 standard candles in the rest-frame near-IR wavelengths, 
 specifically the J and H passbands \citep{barone}. 
 At $z \approx 0.3$ these bands
 redshift into observed H and $\rm K_s$ respectively, so that 
 redshift is a sweet-spot which minimises k-corrections. 

 We note here that this is significantly different to the more 
  common case of computing dark energy figures of merit; in the
  dark energy case, 
   breaking degeneracies between $\Omega_m, \Omega_k$ and
  dark energy parameters $w_0, \, w_a$ 
  requires relative distance measurements spanning a broad 
  range of redshift $0.2 \simlt z \simlt 1.5$; for supernovae
  anchored to local samples at $z \sim 0.05$, SNe at higher
  redshift have greater leverage on $w_0$ and especially $w_a$.   
 Since BAOs are anchored in the CMB, the preference
  for higher redshift is weaker than for SNe, 
  but the rapid increase in available cosmic volume  
 still favours redshifts $0.5 \simlt z \simlt 1.5$ for precision
 measurements of $w_0$ and $w_a$   \citep{wein12}.   

 In contrast to the above, for an absolute $r_s$ measurement 
  we only need to measure an absolute distance $D_L$  
  to one specific redshift matched to a given BAO survey:  
  the overall $r_s$ accuracy is simply the quadrature
   sum of the BAO and $D_L$ errors, with a small addition from 
  the error in Eq.~\ref{eq:dvapp2}, but there is no lever-arm gain
   towards higher redshift.    
  Thus, the number of required SNe for given 
  precision on $D_L$ is independent of the target redshift;
  thus low redshifts are both observationally cheaper, 
 and more robust against systematics such as time evolution
  and imperfect $k-$corrections.  

\subsection {Physical distance measurements} 
\label{sec:physd} 

One major benefit of our approximation \ref{eq:dvapp2} is that
 there is {\em no explicit dependence} on $H_0$.  Therefore, if
  we can measure $D_L(z)$ to $z \sim 0.25$ 
 using some physical-based method which does not rely on calibration 
 via the local distance ladder and $H_0$, we
 automatically bypass the uncertainties in the local
 distance scale.  

There are several current or proposed methods for doing this, 
 including gravitational lens time-delays, 
  Sunyaev-Zeldovich measurements of galaxy clusters, 
 and the expanding-photosphere method applied to Type-II supernovae; 
  however, all of these methods have some level of model dependence
 and it is not yet clear whether they can reach the percent level 
 absolute accuracy (e.g. 2~percent accuracy for a $3\sigma$ detection of
 one additional neutrino species).  
 The lens time-delay method is especially clear at low lens redshift; 
 while lensing observables involve a combination of lens
  and source distances 
 $D_l, D_{ls}$ and $D_s$, 
 selecting systems with $z_l \ll z_s$ makes the ratio $D_{ls}/D_s$
 close to unity and well constrained, which is favourable for
  absolute measurement of the lens distance.  

 Potentially the ultimate $D_L$ calibration 
 method is the detection of gravitational waves
 from coalescing compact binaries \citep{schutz86}, since the model
 waveforms can be predicted extremely precisely assuming only Einstein
 gravity, and the method is completely immune to dust extinction
  or astrophysical nuisance parameters. 
 Of course, such events have not been directly observed so far, but
  the observations of binary pulsars \citep{d-pulsar} leave no doubt that
  the waves exist, and there are ongoing upgrades to Advanced LIGO
 and VIRGO which should give a near-certain detection of 
 binary inspirals around 2015, assuming they reach their design sensitivity. 
 
 These second-generation GW experiments will probably provide
  only modest $D_L$ accuracy for most events; 
  however, if we are lucky there may be a few ``golden events'' 
  with high signal to noise, such as massive 
  black hole events at $z \sim 0.1$.  
 In the longer term, there is an ongoing design study for a 
 third-generation ground-based gravitational wave 
 observatory called ``Einstein Telescope'' 
  \citep{einstein-tel} for the post--2025 era;  this
  is projected to detect binary neutron-star mergers 
 to $z \sim 2$, and neutron-star + black hole mergers to $z \sim 4$.
 For the closest merger events at $z \sim 0.1 - 0.2$, 
  Einstein Telescope would provide very high SNR and percent-level absolute 
 accuracy on $D_L$ for each event.  
  If these can be tied to a unique galaxy, or statistically
  tied to a given cluster or sheet of galaxies, redshift constraints 
  will be quite precise. 

 The future of GW distance measurements is naturally quite uncertain:
 however, one feature is generic: since the method is largely
 limited by instrumental SNR not astrophysical scatter, the
  closest GW inspirals should always provide the best distance precision
  per event.  
 Furthermore, the closer inspirals lead to much smaller
  position error ellipsoids, and make it much easier to
  identify an optical counterpart, or 
 statistically identify the host galaxy in a group or cluster. 
 (Assuming the relative distance and angular errors for a GW inspiral 
   scale $\propto$ 1/SNR, then the comoving volume of the GW error box 
   scales approximately 
   as $D^6$;  this results in fewer candidate 
   host galaxies per burst at low redshift, by a very steep factor).  

 From the above discussion, 
  it is quite generic that for any cosmological 
  distance estimate,
 the best prospects for percent-level absolute accuracy on $D_L(z)$  
 tend to occur at modest redshift $0.1 \simlt z \simlt 0.25$: 
  this is distant enough for galaxy peculiar velocities to be a 
 small effect, but close enough to give high signal/noise ratio and 
   minimal nuisances from possible time evolution and 
  uncorrectable gravitational lensing effects.  
  Until the distant future when we can get cosmological distance
  measurements with significantly better than 1~percent absolute accuracy, 
 then a low redshift will be preferred for anchoring the 
  absolute BAO length.

\subsection{Planck measurement of $r_s$} 

  In the near future, Planck data is expected to
  improve the precision on $\zeq$ to around 1~percent;  
  assuming all the ``standard'' early 
   universe conditions (i.e. GR, standard radiation with
  $\neff = 3.046$, negligible early dark energy, etc),  
  this will determine the sound horizon to $\sim 0.3\,$ percent precision, 
   which is significantly better than any foreseen direct 
  distance measurement. So, why bother measuring $r_s$ locally ?  

  If instead $\neff$ is treated as free, 
  Planck will still measure $r_s \sqrt{\xrad}$ to 0.3~percent, but 
  the error on $\xrad$ will dominate: 
  the Planck measurements of the CMB damping tail (peaks 4,5,6) 
   will provide a useful constraint on $\neff$, 
   but a plausible uncertainty of $\approx 0.3$ in $\neff$ from Planck 
   is equivalent to $4$ percent in $\xrad$  and $2$ percent in $r_s$;
   this is around $6 \times$ worse than the standard-radiation case,    
   and moderate-redshift BAO and distance measurements can potentially be 
   competitive or better than this accuracy.   

  Furthermore, the fitting of the radiation density from 
  the CMB relies on fairly subtle 
  and smooth suppression of power in the CMB damping 
    tail \citep{bs03, hou12}; 
  this effect is significantly degenerate with
  other possible adjustable parameters, including changes 
  in primordial Helium 
  abundance $Y_{p}$ and non-zero running of the primordial spectral index 
  $dn_S / d\ln k$ \citep{hou12}, 
  (and also with possible experimental systematics such as imperfect 
    modelling of beam sidelobes). 
  In CMB analyses, $\neff$, $Y_p$  and $dn_S / d\ln k$ are generally 
  fitted one-at-a-time with 
  the other two fixed to ``standard'' values;  
  however, if two or three of these  
  are simultaneously free, the CMB-only constraints 
  on $r_s(z_d)$ may well be significantly worse than $2\,$ percent; 
  while the local BAO route above can provide a direct measure of $r_s$ 
  which is practically theory--independent.  
 
 Therefore, although cosmic variance means that local BAO measurements
  cannot compete with the 0.3~percent best-case Planck precision on $r_s$, 
   this is not a major drawback:   a local
  measurement of $r_s$ to 1--2 percent absolute accuracy 
   would still be of major benefit for cosmology,
  and could detect or exclude various
  early-universe effects such as non-standard $\neff$ 
   with high significance; this method is 
  independent of early-universe  
  uncertainties including $Y_p$ and spectral
  index running which may potentially hamper the Planck  
  measurement of $\neff$.

 Another motivation is that the value of $\neff$ from the CMB 
  is somewhat degenerate
  with the primordial spectral index $n_s$ and $d n_s/d \ln k$ 
  (e.g. \citealt{hou12}):  
  this can have major implications for constraining 
 the early universe and inflation theory. 
 If $\neff$ is fixed to 4.04 rather than the standard value
  3.04, the WMAP best-fit value of $n_s$ moves up from $\approx 0.96$ 
  to $\approx 0.975$ to compensate; this is
  a small change, but is potentially very important because the 
  scale-invariant value of 1.00 is then no longer ruled out at 
   high significance.
  A constraint on $\neff$ directly from the absolute length 
  $r_s$ is almost independent 
  of the primordial power spectrum, and is therefore extremely valuable.

In principle we can achieve better precision by going to a higher 
  redshift BAO survey to reduce cosmic variance,  
 e.g. Euclid should measure the transverse BAO angle 
  $r_s / D_A(z)$ to better than 0.4 percent in many bins between
    $0.7 \le z \le 1.7$ \citep{euclid-red}. 
 Adding a 0.4 percent distance measurement to a matching redshift,  
  this could measure $\neff$ to around $\pm 0.1$ precision,
   which is substantially better than Planck. 
  However, a sub--percent absolute distance to such a redshift
   currently appears extremely challenging given the potential systematics: 
 thus the low-redshift route outlined above remains a 
 promising intermediate step.

\subsection{An ultimate BAO survey at z $\sim$ 0.2} 

 The considerations above on distance measurements and CMB degeneracies 
 provide a very strong motivation for obtaining the best possible
  BAO measurements at modest $z \sim 0.2$, approaching 
  the cosmic variance limit. 
  The ongoing BOSS project is
 a large step in this direction, but there are several potential
 improvements: firstly of course BOSS only covers around
  1/4 of the entire sky, so adding coverage of the 
 Southern hemisphere is very useful; secondly, sampling a somewhat higher 
 space density of galaxies can improve reconstruction of the BAO peak, 
  and thirdly we may expand the survey to lower galactic latitudes for 
  maximal sky coverage. 

 Until recently, galaxy surveys have disfavoured low galactic latitudes
  due to both extinction problems and increased stellar contamination  
 (e.g. from blended images which are hard to morphologically classify). 
  However, the recently completed WISE mid-IR survey 
   combined with the ongoing VISTA Hemisphere Survey
  should provide a galaxy sample of ample depth, and minimal 
  sensitivity to galactic extinction which could push down to
  $\vert b \vert \sim 15^o$. Availability of optical+near-IR colours 
 can also greatly improve the star-galaxy separation, so 
 stellar contamination should remain manageable.   
  The cosmic-variance limits on BAO measurements have been calculated
  by \citet{se07}; for 3/4 of the full sky and realistic reconstruction
  methods, interpolation from their Figure~3 predicts precision 
  $\approx 1.2$ percent on $r_s / D_V(z = 0.2)$; this accuracy is similar to
  optimistic projections for local $H_0$ measurements. 
  Such a BAO survey is comfortably 
  within reach of proposed high-multiplex multi-object spectrographs 
  such as 4MOST on the VISTA telescope, or DESpec at CTIO. 
  The required area is very large, but the target density $\sim 50$
   galaxies per deg$^2$ is rather low, so
  such an observing program would only take a modest fraction
  of the total number of fibres, and could be run in a simultaneous
  mode in parallel with stellar and other surveys.  

 Furthermore, an accurate low-redshift BAO measurement, when compared 
   to a radial BAO
  measurement at $z \sim 0.7$, can provide a clean smoking--gun test
  of cosmic acceleration entirely from the two BAO measurements 
  \citep{suth12}; that test 
   is independent of supernovae, CMB data and general relativity.
  BAO measurements at $z \simgt 0.5$ are necessary but 
  not sufficient for
  this test, since very little acceleration happened earlier 
  than $z = 0.5$.

\section{Conclusion} 
\label{sec:conc} 

 Measuring the absolute rather than relative BAO length scale  
 forms a powerful test of standard early-universe cosmology, especially
  probing the radiation density along with other possible non-standard 
 effects at $z > 1000$.  

 As a step in this direction, we have found a simple and highly accurate 
  approximation (Eq.~\ref{eq:dvapp2}) 
 relating the BAO dilation scale $D_V(z)$ to the luminosity distance
 $D_L$ at a slightly higher redshift. This is accurate to $\le 0.2$ percent
 at $z \le 0.4$ for all plausible WMAP-compatible Friedmann models,
  including modest curvature and time-varying dark energy; 
  the inaccuracy is substantially
  smaller than the cosmic variance limit for low-redshift BAO
  measurements.     The approximation does
  not explicitly depend on $H_0$, so remains applicable if
  there is any direct physics-based measurement of $D_L(z)$ 
 bypassing the local distance ladder.  
 The only ways for Eq.~\ref{eq:dvapp2} to have percent-level errors
  are very radical, 
  such as violation of the distance-duality relation $D_L = (1+z)^2 D_A$, 
  or a sharp phase transition in dark energy 
  at low redshift, e.g. a sharp jump in $w(z)$ causing a 
  kink feature in $H(z)$.  

 We also reviewed the degeneracy between radiation density and 
 cosmic scales, and showed this is close to a rescaling of all dimensionful
  observables (except baryon and photon densities), 
  while leaving most dimensionless ratios unchanged.  

 Given realistic future observations, 
  the approximation above can provide a high-precision calibration 
 of the BAO length scale using only low redshift data, 
  which in turn provides a powerful test of standard 
  $z > 1000$ CMB assumptions, and in particular a robust test of
  the radiation density independent of the CMB damping tail.   

  A measurement of $H_0$ is also useful, but on its own does not
   fully break degeneracies: e.g. a high-precision measurement of $H_0$
 significantly greater than $73 \hunit$ would signal a problem 
  for vanilla $\Lambda$CDM, but could indicate any one of 
   $w < -1$, weak open curvature or increased radiation density, and 
  without an absolute $r_s$ measurement it would be hard 
  to discriminate these. 
  In contrast, an 
  absolute BAO length measurement can cleanly detect or constrain 
  non-standard pre-CMB physics, almost independent of late-time effects 
  such as $w \ne -1$ or weak curvature, and with minimal degeneracy with
  $n_s$ and running spectral index.   
  This may also be important for inflation theory, since the currently
  strong evidence for $n_s < 1$ becomes substantially 
   weaker if $\neff$ is larger than the standard value.  

 We can essentially distinguish two possibilities:
  if all the standard CMB assumptions are correct, then Planck
  will determine $r_s(z_d)$ better than the cosmic variance
  on the BAO length: then an absolute BAO length measurement
  essentially provides a strong null test of the standard 
  cosmology at around 1-2 percent precision, but does not improve our
   error bars on $\Omega_m$, $w$ etc.  

  However, if one or more of the standard early-universe assumptions is
   wrong, this can be absorbed into biased values of $H_0$, 
  and to a lesser extent $\Omega_m$
   and $w$, in joint fits to CMB, BAO, and supernova data alone.   
  Therefore,  a direct low-redshift measurement of $r_s$
  can be very powerful for discriminating
   early-universe modifications such as extra radiation 
  or early dark energy, from late-time effects 
  such as dark energy $w \ne -1$ or small non-zero curvature.

\section*{Acknowledgements}

I thank Will Percival for helpful discussions, and Roelof de Jong 
  and the 4MOST science team for information on survey strategies. 
I acknowledge the use of WMAP data from 
 the Legacy Archive for Microwave Background
 Data Analysis (LAMBDA) at GSFC (lambda.gsfc.nasa.gov),
  supported by the NASA Office of Space Science. 

\vspace{5mm} 
(The definitive version of this paper is available in MNRAS at 
  DOI:10.1111/j.1365-2966.2012.21666.x ) 

\clearpage 
 

\onecolumn 
\appendix

\section[]{The Taylor series for $D_V$ }

 Here we derive the Taylor series for $D_V(z)$ and 
 approximation~\ref{eq:dvapp1}; {\newtwo these are not used in the main 
 part of the paper, but are useful } 
 to provide an analytic understanding of the high accuracy of approximations
 (\ref{eq:dvapp1}) and (\ref{eq:dvapp2}), {\newtwo and the 
 dependence on cosmological parameters.}

We start by defining the usual 
 deceleration parameter $q$ and the jerk parameter $j$ 
 (e.g. \citealt{alam03}) as  
\begin{equation} 
 q \equiv -{d^2 a/dt^2  \over a H^2} \ , 
  \qquad j \equiv { d^3a / dt^3 \over a H^3} \ . 
\end{equation} 

We can rearrange these in terms of $d/dz$, and using the chain rule
 we find 
\begin{equation} 
\label{eq:ddhinv} 
 {d \over dz} \left( \frac{1}{H} \right) =  {1 + q \over (1+z) H} 
  \qquad , \qquad \qquad  
  {d^2 \over dz^2}\left( \frac{1}{H} \right) = 
   { 2 + 4q + 3q^2 - j \over (1+z)^2 \,H }  \ . 
\end{equation} 

Using these, with subscript $0$ denoting present-day values, 
  we obtain the series 
\begin{equation}
 {c \, z \over H(z)} = {c \, z \over H_0} \left[ 1 - (1 + q_0) z 
  + {2 + 4q_0 + 3q_0^2 - j_0 \over 2} z^2 + \ldots \right] 
\end{equation}
 and integrating and including the leading-order curvature term gives
\begin{equation}
\label{eq:datay} 
 (1+z) D_A(z) \; = \;  
 {c \, z \over H_0} \left[ 1 - {1 + q_0 \over 2} z 
   +   {2 + 4q_0 + 3q_0^2 - j_0 + \Omega_k \over 6} \, 
  z^2 + \ldots \right] 
\end{equation}
Inserting the above two expressions 
 into the definition of $D_V$, collecting powers of $z$ 
 and using $(1+x)^{1/3} = 1 + x/3 - x^2/9 + \ldots$, 
 we obtain the Taylor series for $D_V$ as 
\begin{equation} 
\label{eq:dvtay} 
D_V(z) \; = \; 
 {c \,z \over H_0} \left[ 1 - {2(1+q_0) \over 3} z  
   +  {19 + 38 q_0 + 29 q_0^2 - 10 j_0 + 4 \Omega_k \over 36} 
  \, z^2 + \ldots \right] 
\end{equation} 
(We note that for concordance $\Lambda$CDM, 
  the 3-term sum above has error $< 0.25\,$ percent at $z < 0.3$, 
 but worsens quite rapidly above this.)  

Substituting $4z/3$ in \ref{eq:datay}, we have  
\begin{equation} 
\label{eq:datay43} 
 \frac{3}{4} \left( 1+ \frac{4}{3}z \right) D_A(\frac{4}{3}z) \; = \; 
  {c \, z \over H_0} \left[ 1 - {2 (1 + q_0) \over 3} z  
  + \, \frac{16}{54} (2 + 4q_0 + 3q_0^2 - j_0 + \Omega_k) \, z^2 
  + \ldots \right] 
\end{equation}

Comparing the above two equations, 
 it is clear that approximation \ref{eq:dvapp1} is correct to second
 order in $z$, for any values of cosmological parameters.    
Subtracting  (\ref{eq:dvtay}) from 
 (\ref{eq:datay43}) and dividing by 
 (\ref{eq:dvtay}), we then find that the ratio of the RHS to LHS of  
  approximation~\ref{eq:dvapp1} is  
\begin{equation} 
   { 3 \over 4}  { (1+ \frac{4}{3}z) D_A(\frac{4}{3}z) 
  \over D_V(z) } \; = \; 
   1 + \frac{z^2}{108} 
  \left[ \frac{14}{9} - 2 j_0 + 9 \left(q_0 + \frac{7}{9}\right)^2 
  + 20 \,\Omega_k \right] + O(z^3) \ . 
\end{equation} 

A flat $\Lambda$CDM model has $q_0 = \frac{3}{2} \Omega_m - 1$ and
 $j_0 = +1$, hence the term in square brackets above 
 simplifies to $\frac{1}{4} \Omega_m (81 \,\Omega_m - 24)$: this is 
 {\newtwo less than 1 for} 
 plausible values of $\Omega_m < 0.4$ (it is
  $-0.16$ for the concordance model).  

More generally, 
 for conservative ranges of parameters $-1 < q_0 < -0.4$, $0 < j_0 < 2$
 and $\vert \Omega_k \vert < 0.05$, the square--bracket 
  is not significantly bigger than $\pm 4$; with the prefactor
 of $1/108$,  this explains the excellent
 accuracy of approximation~\ref{eq:dvapp1} at moderate redshift. 
{\newtwo  This also suggests that approximation~\ref{eq:dvapp1} 
 should remain fairly 
  accurate for modified-gravity models, as long as they are homogeneous, have
  weak curvature and $q_0, \; j_0$ not very different from the 
 concordance model. }  

 Finally we note that the square--bracket 
  has value 14.25 for an Einstein-de Sitter model, 
 and approximately 22 for a zero-$\Lambda$ open model, explaining the
 low-redshift limit of those models shown in Figure~\ref{fig:dvapp1}. 

 \newpage

\section{Large-scale structure and $\zeq$}
\label{sec:appb}

Here we note that while BAO parameter estimates 
 are strongly dependent
 on correct deduction of $\zeq$ from the CMB, the
 overall shape of the large-scale galaxy power spectrum does provide
  an independent check of this. 

 The large-scale galaxy clustering pattern 
 actually contains two key length scales, the BAO length discussed above,
 and also the ``big bend'' scale which describes the overall broad-band shape
 of the galaxy power spectrum $P(k)$ excluding the BAO wiggles;  
   these two are approximately independent observables. 
  Assuming the primordial power 
  spectrum is well described by a power-law, $n_S \sim 0.96$ and 
  the dark matter is cold or warm (not hot), then fitting
  the ``big bend'' in a galaxy power spectrum 
  essentially measures the comoving light horizon 
  size $r_H$ at matter-radiation equality, again relative to 
   $D_V(z)$ at the characteristic redshift of the given survey: this gives
\begin{equation} 
\label{eq:rh} 
 z \, { r_H (\zeq) \over D_V(z) } =  (1 + \epsilon_V) 
   { E(2z/3) \over \sqrt{\Omega_m} } { 2 (\sqrt{2} - 1) \over 
    \sqrt(1 + \zeq)  } \ ; 
\end{equation} 
  which is not explicitly dependent on $\xrad$. 
 For $\xrad = 1$, at small $z$ the above has the well-known scaling   
  $\propto (\Omega_m h)^{-1}$, with a result 
  $\Omega_m h \approx 0.20$ which has remained consistent 
  over many large galaxy surveys, since the first reliable estimate 
  from the APM Galaxy Survey \citep{esm90}, through 2dFGRS 
   \citep{psp02} and SDSS \citep{reid10}.  

 We see that the big-bend observable in (\ref{eq:rh})  
  has the same $E(2z/3)/\sqrt{\Omega_m}$ factor as the BAO ratio 
  in (\ref{eq:omez}),
   but has a different dependence on $\zeq$.
  Taking the ratio of these two characteristic lengths, we
   have 
 \begin{eqnarray}  
 \label{eq:rsrh} 
 {r_s(z_d) \over r_H(\zeq) } & \equiv & { r_H(z_d) \over r_H(\zeq) } 
  \, { r_s(z_d) \over r_H(z_d) } \nonumber \\ 
    & \simeq &  { \left(  {1+\zeq \over 1+z_d} + 1 \right)^{1/2} \ - 1 
    \over \sqrt{2} - 1 } \;  { 0.886 \over \sqrt{3} }  \ ;  
\end{eqnarray} 
  here the second factor $0.886 / \sqrt{3}$ represents 
  the weighted average sound speed $c_s/c$ prior to the drag redshift; 
 the given value is for the concordance model, but 
 this term is very insensitive to reasonable parameter variations. 
 The first term above depends only on the ratio 
  $(1 + \zeq)/(1 + z_d)$, and scales 
   approximately $\propto (1+\zeq)^{0.75}$ around
  the concordance model;  there is no
  separate dependence on $\Omega_m$, $h$ and $\xrad$, so this ratio
   is predicted robustly given just $\zeq$ from WMAP alone.   

 Due to various uncertainties in overall $P(k)$ shape 
  from possible effects such as scale-dependent bias, non-linearity,
   redshift-space distortions, 
  neutrino masses, running of $n_s$ etc, 
  this BAO/bend ratio seems unlikely to independently 
  measure $\zeq$ to a precision 
   comparable to the current 4~percent precision from WMAP, 
  and still less the 1~percent expected from Planck.  
  However, the fact that parameters 
   estimated from CMB+BAO also provide a reasonable fit 
   to the overall $P(k)$ shape provides a valuable consistency check, 
  which the concordance model passes \citep{reid10}. 
  This strongly argues that the WMAP-only estimate of $\zeq$ {\em cannot} 
   have a gross error from unknown physics, 
  unless there has been a fortuitous cancellation of effects. 

  The above also shows that measuring $P(k\, r_s)$ in dimensionless units
   rescaled by the BAO length may be helpful for 
  testing for neutrino masses or non-standard physics 
  around $z \sim \zeq$, since this is very robust against shifts
  in $\Omega_m$, $H_0$ and $\xrad$.  

\bsp

\label{lastpage}


\begin{thebibliography}{99}

%
\bibitem[\protect\citeauthoryear{Ade et al}{2011}]{planck-miss} 
  Ade, P.A.R. et al (Planck Collaboration), 2011, A\&A, 536, 1. 
\bibitem[\protect\citeauthoryear{Alam et al}{2003}]{alam03}
  Alam U., Sahni V., Saini T-D., Starobinsky A.A., 2003,
  {MNRAS}, 344, 1057.
\bibitem[\protect\citeauthoryear{Barone-Nugent et al}{2012}]{barone} 
  Barone-Nugent R.L. et al, 2012, arXiv.org/1204.2308 
\bibitem[\protect\citeauthoryear{Bashinsky \& Seljak}{2004}]{bs03} 
  Bashinsky S. \& Seljak U., 2004, Phys. Rev. D, 69, 083002.  
\bibitem[\protect\citeauthoryear{Bassett \& Hlozek}{2010}]{bh10} 
 Bassett B.A. \& Hlozek R., 2010, in ``Dark Energy'', ed P. Ruiz-Lapuente, 
 p246, Cambridge Univ. Press, Cambridge.  
\bibitem[\protect\citeauthoryear{Bernstein et al}{2012}]{des-sn} 
 Bernstein J.P., Kessler R., Kuhlmann S. et al 2012, ApJ, 753, 152.
\bibitem[\protect\citeauthoryear{Beutler et al}{2011}]{beutler11} 
 Beutler F., Blake C., Colless M. et al, 2011, MNRAS, 416, 3017. 
\bibitem[\protect\citeauthoryear{Blake \& Glazebrook}{2003}]{bg03} 
 Blake C. \& Glazebrook K., 2003, ApJ, 594, 665.  
\bibitem[\protect\citeauthoryear{Blake et al}{2011}]{blake11} 
 Blake C., Davis T., Poole G. et al, 2011, MNRAS, 415, 2892. 
\bibitem[\protect\citeauthoryear{Bond \& Efstathiou}{1984}]{be84} 
 Bond J.R., Efstathiou G., 1984, ApJ, 285, L45 
%
\bibitem[\protect\citeauthoryear{Cole et al}{2005}]{cole05} 
  Cole S., Percival  W.J., Peacock J.A. et al, 2005, { MNRAS}, 362, 505.  
\bibitem[\protect\citeauthoryear{de Bernardis et al}{2008}]{debernardis} 
  de Bernardis F., Melchiorri A., Verde L., Jimenez R., 2008, JCAP, 03, 020. 
\bibitem[\protect\citeauthoryear{Efstathiou \& Bond}{1999}]{eb99} 
 Efstathiou G., Bond J.R., 1999, MNRAS, 304, 75.  
\bibitem[\protect\citeauthoryear{Efstathiou, Sutherland \& Maddox}
  {1990}]{esm90} 
 Efstathiou G., Sutherland W., Maddox S., 1990, Nature, 348, 705. 
\bibitem[\protect\citeauthoryear{Eisenstein \& Hu}{1998}]{eh98} 
 Eisenstein D.J. \& Hu W., 1998, { ApJ}, 496, 605. 
\bibitem[\protect\citeauthoryear{Eisenstein \& White}{2004}]{ew04} 
 Eisenstein D.J., White M., 2004, Phys.Rev.D, 70, 103523. 
\bibitem[\protect\citeauthoryear{Eisenstein et al}{2005}]{eis05} 
 Eisenstein D.J., Zehavi I., Hogg D. et al, 2005, { ApJ}, 633, 560.  
\bibitem[\protect\citeauthoryear{Eisenstein et al}{2007a}]{esss07} 
 Eisenstein D.J, Seo H., Sirko E., Spergel D.N., 2007, { ApJ}, 664, 675.  
\bibitem[\protect\citeauthoryear{Eisenstein, Seo \& White}{2007b}]{esw07} 
 Eisenstein D.J, Seo H., White M., 2007, { ApJ}, 664, 660.  
\bibitem[\protect\citeauthoryear{Hou et al}{2012}]{hou12} 
Hou Z., Keisler R., Knox L., Millea M., Reichardt C.,
   2011, arXiv.org/1104.2333  
\bibitem[\protect\citeauthoryear{Kaiser et al}{2010}]{pan-starrs} 
 Kaiser N., Burgett W., Chambers K. et al, 2010, Proc. SPIE, 7733, 12. 
\bibitem[\protect\citeauthoryear{Keisler et al}{2011}]{keis11} 
  Keisler R., Reichardt C.L., Aird K.A. et al, 2011, ApJ, 743, 28. 
\bibitem[\protect\citeauthoryear{Komatsu et al}{2011}]{komatsu11} 
  Komatsu E., Smith K., Dunkley J. et al, 2011, { ApJS}, 192, 18. 
\bibitem[\protect\citeauthoryear{Kramer \& Stairs}{2008}]{d-pulsar} 
 Kramer M., Stairs I.H., 2008, ARA\&A, 46, 541. 
\bibitem[\protect\citeauthoryear{Laureijs et al}{2011}]{euclid-red} 
  Laureijs, R. et al, 2011, Euclid Red Book, arXiv.org/abs/1110.3193 
\bibitem[\protect\citeauthoryear{Linder \& Robbers}{2008}]{lin-rob} 
   Linder E., Robbers G., 2008, JCAP, 06, 004.   
\bibitem[\protect\citeauthoryear{Mangano et al}{2005}]{mangano05} 
 Mangano G., Miele G., Pastor S., Pinto T., Pisanti O., Serpico P.D., 
   2005, Nucl. Phys. B, 729, 221. 
\bibitem[\protect\citeauthoryear{Mangano \& Serpico}{2011}]{mangano11} 
 Mangano G., Serpico P.D.,  2011, Phys. Lett. B, 701, 296. 
\bibitem[\protect\citeauthoryear{Meiksin, White \& Peacock}{1999}]{mwp99} 
 Meiksin A., White M. \& Peacock J.A., 1999, { MNRAS}, 304, 851. 
\bibitem[\protect\citeauthoryear{Menegoni et al}{2012}]{menegoni} 
  Menegoni E., Archidiacono M., Calabrese E. et al, 2012, 
  Phys. Rev. D, 85, 107301. 
\bibitem[\protect\citeauthoryear{Peebles \& Yu}{1970}]{peeb-yu70} 
 Peebles P.J.E. \& Yu J.T, 1970, ApJ, 162, 815. 
\bibitem[\protect\citeauthoryear{Padmanabhan et al}{2012}]{pad12} 
  Padmanabhan N., Xu X., Eisenstein D.J., Scalzo R., Cuesta A.J., 
   Mehta K.T., Kazin E., 2012, arXiv.org/1202.0090 
\bibitem[\protect\citeauthoryear{Percival et al}{2002}]{psp02} 
  Percival W.J., Sutherland W., Peacock J.A. et al, 2002, MNRAS, 337, 1068. 
\bibitem[\protect\citeauthoryear{Percival et al}{2010}]{perc10} 
  Percival W.J., Reid  B.A., Eisenstein D.J. et al, 2010, MNRAS, 401, 2148. 
\bibitem[\protect\citeauthoryear{Reid et al}{2010}]{reid10} 
 Reid  B.A.,  Percival W.J., Eisenstein D.J. et al, 2010, MNRAS, 404, 60. 
\bibitem[\protect\citeauthoryear{Sathyaprakash et al}{2011}]{einstein-tel} 
 Sathyaprakash B., Abernathy M., Acernese F. et al, 2011, 
  arXiv.org/1108.1423 
\bibitem[\protect\citeauthoryear{Schutz}{1986}]{schutz86} 
 Schutz B.F., 1986, Nature, 323, 310. 
\bibitem[\protect\citeauthoryear{Seo \& Eisenstein}{2003}]{se03} 
 Seo H-J., Eisenstein D.J., 2003, ApJ, 598, 720. 
\bibitem[\protect\citeauthoryear{Seo \& Eisenstein}{2007}]{se07} 
 Seo H-J., Eisenstein D.J., 2007, ApJ, 665, 14. 
\bibitem[\protect\citeauthoryear{Seo et al}{2012}]{seo12} 
 Seo H-J. et al, 2012, arXiv.org/1201.2172. 
\bibitem[\protect\citeauthoryear{Seo et al}{2010}]{seo10} 
 Seo H-J., Eckel J., Eisenstein D.J. et al,  2010, ApJ, 720, 1650. 
\bibitem[\protect\citeauthoryear{Seo et al}{2008}]{seo08} 
 Seo H-J., Siegel E.R., Eisenstein D.J., White M.,  2008, ApJ, 686, 13. 
\bibitem[\protect\citeauthoryear{Stern et al}{2010}]{stern10} 
 Stern D., Jimenez R., Verde L., Kamionkowski M., 
  Stanford S.A., 2010, JCAP, 02, 008. 
\bibitem[\protect\citeauthoryear{Sutherland}{2012}]{suth12} 
 Sutherland W., 2012, MNRAS, 420, 3026.  (arXiv.org/abs/1105.3838)  
\bibitem[\protect\citeauthoryear{Weinberg et al}{2012}]{wein12} 
 Weinberg D.H., Mortonson M.J., Eisenstein D.J., Hirata C., Reiss A.G., 
  Rozo E., 2012, Phys. Reports, submitted (arXiv.org/1201.2434)   
\bibitem[\protect\citeauthoryear{White et al}{2011}]{white11} 
 White M., Blanton M., Bolton A. et al, 2011, ApJ, 728, 126.  
\bibitem[\protect\citeauthoryear{Zunckel et al}{2011}]{zunckel} 
 Zunckel C., Okouma P., Muya Kasanda S., Moodley K, Bassett B.A.,
  2011, Phys.Lett.B, 696, 433. 

\end{thebibliography}
\end{document}